\documentclass{article}
\usepackage{arxiv}
\usepackage[utf8]{inputenc}
\usepackage[T1]{fontenc}
\usepackage{xcolor}
\usepackage[round]{natbib}
\definecolor{blendedblue}{rgb}{0.2, 0.2, 0.6}
\usepackage{graphicx}
\usepackage[bookmarks   = true,
            citecolor   = blendedblue,
            colorlinks  = true,
            linkcolor   = blendedblue,
            urlcolor    = blendedblue,
            citecolor   = blendedblue,
            linktocpage = false,
            hyperindex  = true]{hyperref}
\usepackage{booktabs}
\usepackage{amsfonts}
\usepackage{enumitem}
\usepackage{nicefrac}
\usepackage{amsmath, amssymb, amsthm}
\usepackage{url}
\usepackage{doi}
\usepackage{bm}
\usepackage{fontawesome}
\usepackage{mathtools}
\usepackage{lmodern}
\usepackage{siunitx}
\usepackage{booktabs}
\usepackage{etoolbox}
\usepackage{calc}
\usepackage{dsfont}
\usepackage{multirow}
\usepackage{array}
\usepackage{graphicx}
\usepackage{todonotes}
\usepackage[linesnumbered,ruled,vlined]{algorithm2e}
\usepackage{orcidlink}

\definecolor{blendedblue}{rgb}{0.2, 0.2, 0.6}
\definecolor{forestgreen(web)}{rgb}{0.13, 0.55, 0.13}
\definecolor{darkorange}{rgb}{1.0, 0.55, 0.0}
\definecolor{emeraldblue}{HTML}{1eb5be}

\usepackage{amsfonts, amsmath, amssymb, enumerate} 

\usepackage[labelsep=period]{caption}

\renewcommand{\headeright}{}

\usepackage{parskip}
\makeatletter 
\renewcommand\@biblabel[1]{#1.} 
\makeatother

\title{Feature Reconstruction and Monitoring of Load Test Data under Varying Environmental Conditions} 

\renewcommand{\shorttitle}{Feature Reconstruction and Monitoring of Load Test Data under Varying Environmental Conditions}

\date{\today}

\author{
    Lizzie Neumann \orcidlink{0000-0003-2256-1127} \\
    Chair of Statistics and Data Science \\
 	Dept. of Mathematics and Statistics\\
	School of Economics and Social Sciences\\
    Helmut Schmidt University\\
	Hamburg, Germany\\
	\texttt{neumannl@hsu-hh.de} \\
 	\And
    Philipp Wittenberg \orcidlink{0000-0001-7151-8243} \\
    Chair of Statistics and Data Science \\
 	Dept. of Mathematics and Statistics\\
	School of Economics and Social Sciences\\
    Helmut Schmidt University\\
	Hamburg, Germany\\
	\texttt{pwitten@hsu-hh.de} \\
	\And
    Alexander Mendler \orcidlink{0000-0002-7492-6194} \\
    Chair of Non-destructive Testing \\
    Dept. of Materials Engineering\\
    TUM School of Engineering and Design\\
	Technical University of Munich\\
    Munich, Germany\\
	\texttt{alexander.mendler@tum.de}\\
	\And 
    Jan Gertheiss \orcidlink{0000-0001-6777-4746} \\
    Chair of Statistics and Data Science \\
 	Dept. of Mathematics and Statistics\\
 	School of Economics and Social Sciences\\
    Helmut Schmidt University\\
	Hamburg, Germany\\
	\texttt{gertheij@hsu-hh.de} \\
}

\begin{document}	
\maketitle

\begin{abstract}
System outputs in Structural Health Monitoring (SHM), such as sensor measurements or extracted features like eigenfrequencies, are influenced not only by (potential) damage but also by environmental and operational variables (EOV). Identifying these factors and removing their effects from the data is essential before proceeding with further analysis. Most existing methods for this task focus on the expected values of system outputs, e.g., using different types of response surface modeling. However, it has been shown that confounding variables can also affect the (co-)variance of and between system outputs. This is particularly important because the covariance matrix is an essential building block in many damage detection methods in SHM. Beyond standard response surface modeling, a nonparametric kernel approach can be used to estimate a conditional covariance matrix that can change depending on the identified confounding factor. This improves our understanding of how, e.g., temperature affects the system outputs.

In this work, we present a new confounder-adjusted version of feature reconstruction. It uses the conditional covariance matrix as the basis for (conditional) principal component analysis. The resulting (conditional) principal component scores are then used to reconstruct system outputs with the confounding influences removed.  In particular, the new approach eliminates the confounder’s effect on both the mean and the covariance. As will be shown on load test data from the Vahrendorfer Stadtweg bridge in Hamburg, Germany, the reconstructed features can then be employed for monitoring, e.g., using an appropriate control chart, resulting in fewer false alarms and a higher probability of detecting damage.
\end{abstract}

\bigskip

\section{Introduction}

In structural health monitoring (SHM), structures such as bridges and buildings are tracked over time using regularly recorded measurements to detect changes in their material and geometric properties. Before anomalies can be detected, the effects of environmental and operational variables (EOVs) have to be removed from system outputs, such as sensor measurements or derived features. 
This process is known as ``data normalization'' \cite{Farrar.Worden_2013} and there are various literature reviews available \cite{Avci.etal_2021,Han.etal_2021,Wang.etal_2022}. If a physically meaningful feature is reconstructed, data normalization is also called ``feature reconstruction''. Data normalization is essential because EOVs can cause false alarms in damage detection. 
There are several data normalization methods that can be divided into two groups: ``supervised'' and ``unsupervised'' methods, depending on whether EOVs must be measured. Unsupervised method use, e.g., various types of principal component analysis (PCA) \cite{Magalhaes2012, Reynders.etal_2014}, auto-associative neural networks \cite{Worden2007}, or non-parametric machine learning~\cite{Entezami.etal_2022}. Supervised methods include, e.g., ``response surface models'' \cite{Doehler.etal_2014}, methods using support vector machines \cite{Ni2005}, or neural networks \cite{Cury2012}. What all these methods have in common is that they focus solely on the expected (mean) values of the system outputs and ignore the EOV's effects on measurement error. But \cite{Neumann.etal_2025} showed that not only the mean but also second-order statistical moments of system outputs need to be adjusted for EOVs; in particular, accounting for changing temperatures is essential. For doing so, \cite{Neumann.etal_2025} developed a conditional version of PCA (thus, turning it into a supervised method). This approach can also be used for reconstructing physically meaningful features that are no longer influenced by temperature in the mean and covariance \cite{Neumann.etal_2025c}. In this paper, the approach is applied to the load test data of the Vahrendorfer Stadtweg bridge.

The remainder of the paper is organized as follows. Section~\ref{vs} presents the Vahrendorfer Stadtweg bridge and data set used for analysis. Section~\ref{sec_CC_FR} discusses conditional principal component analysis (cPCA) using conditional covariances, confounder-adjusted feature reconstruction, and control charts for monitoring. Section~\ref{AppSHMData} applies the proposed method to load test data from the Vahrendorfer Stadtweg bridge and compares it to two existing methods. The analysis shows that the new method outperforms existing methods and reliably detects additional masses placed on the bridge, whereas the other methods fail to do so. Section~\ref{conclusion} concludes the paper. 

\section{Load Tests on the Vahrendorfer Stadtweg Bridge}\label{vs}

In this paper, conditional PCA-based feature reconstruction will be applied to data from a real-world bridge, the ``Vahrendorfer Stadtweg'' bridge, where strain measurements will be analyzed. 
The outputs will be reconstructed for both in-control and out-of-control data to demonstrate that the temperature's confounding effect is removed, but the ``damage'' is still detectable. 

\begin{figure}[ht]
    \centering
    \includegraphics[height = .34\textwidth]{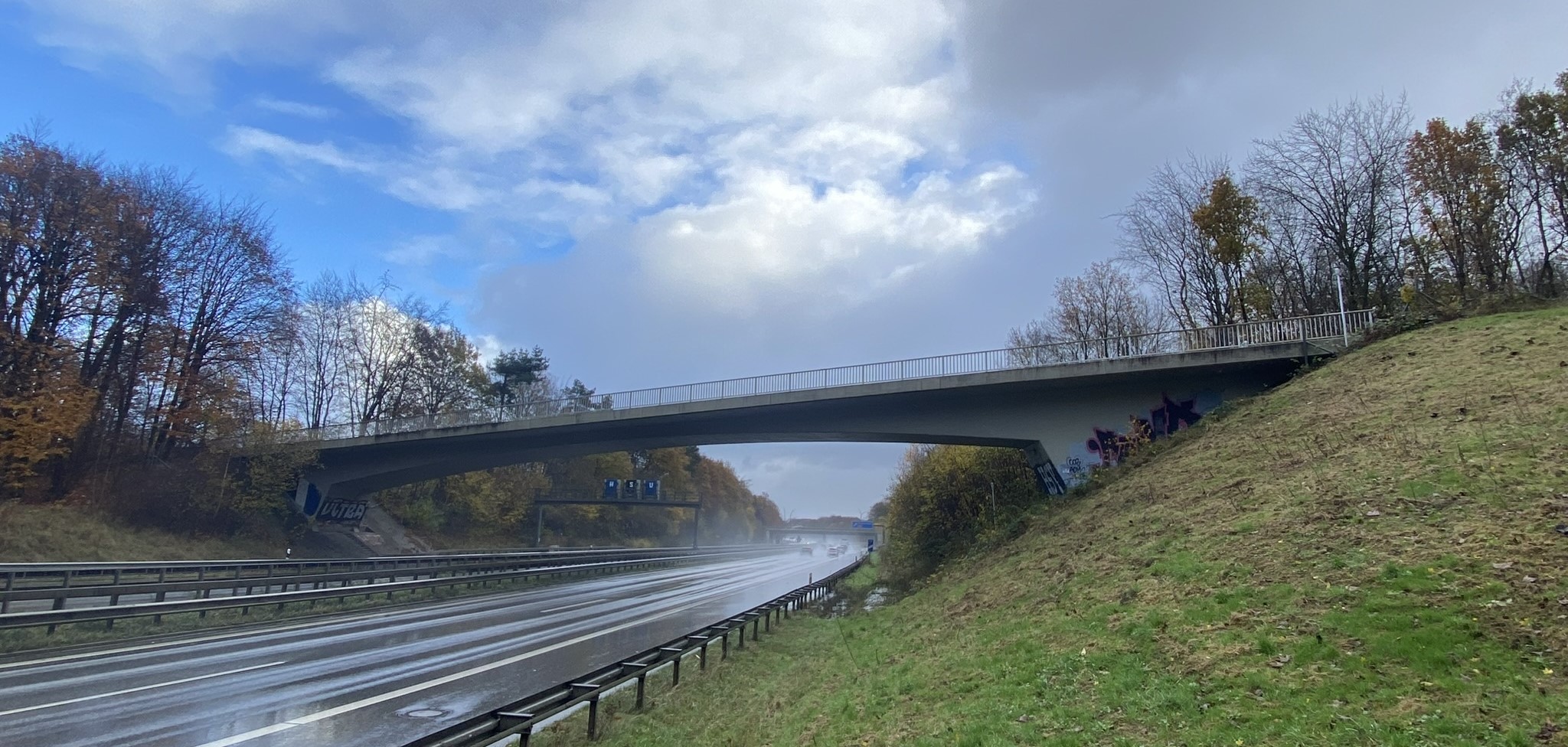}
    \includegraphics[height = .34\textwidth]{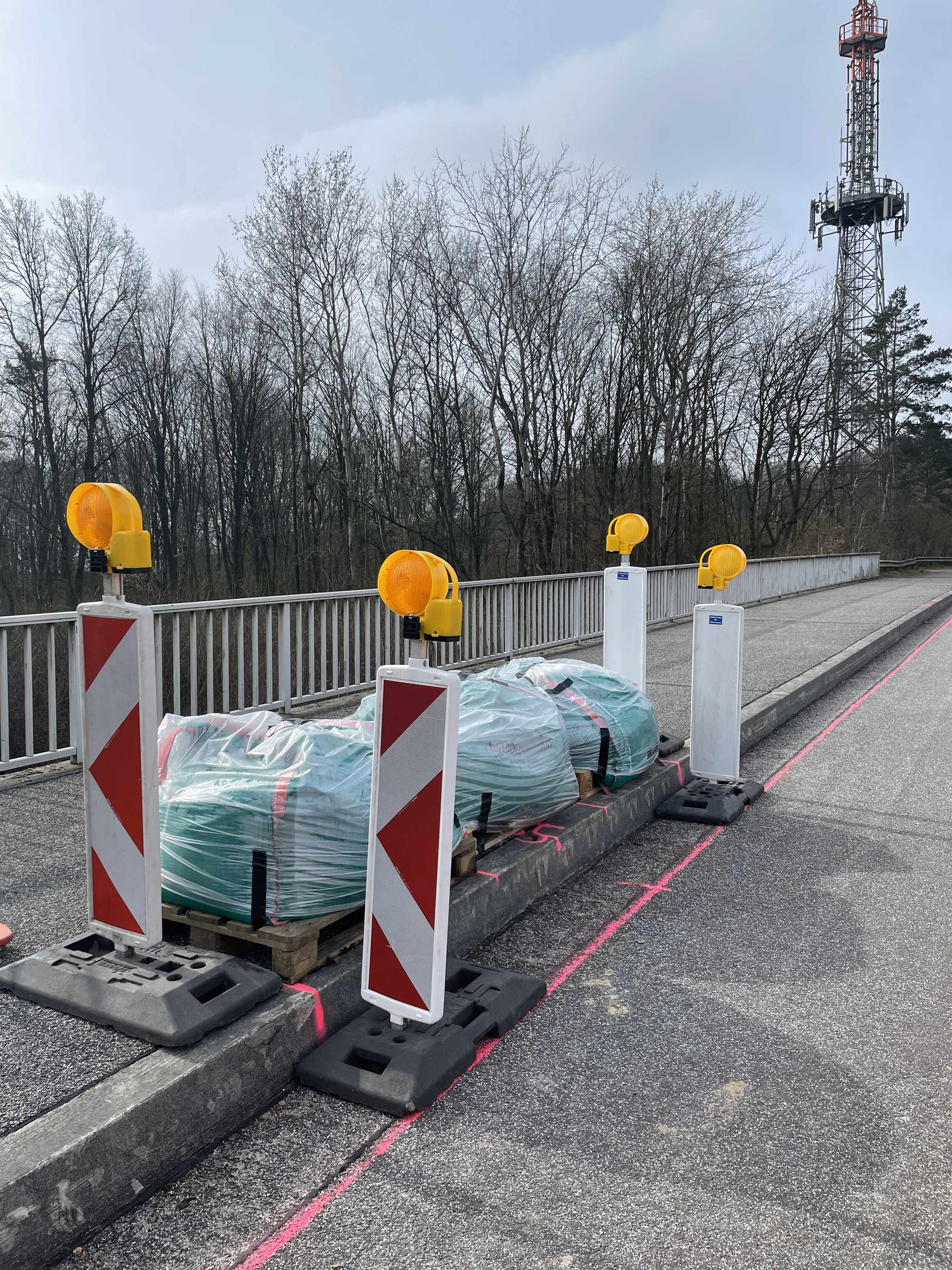}\\ 
    \vspace{.2mm}
    \includegraphics[width=\linewidth]{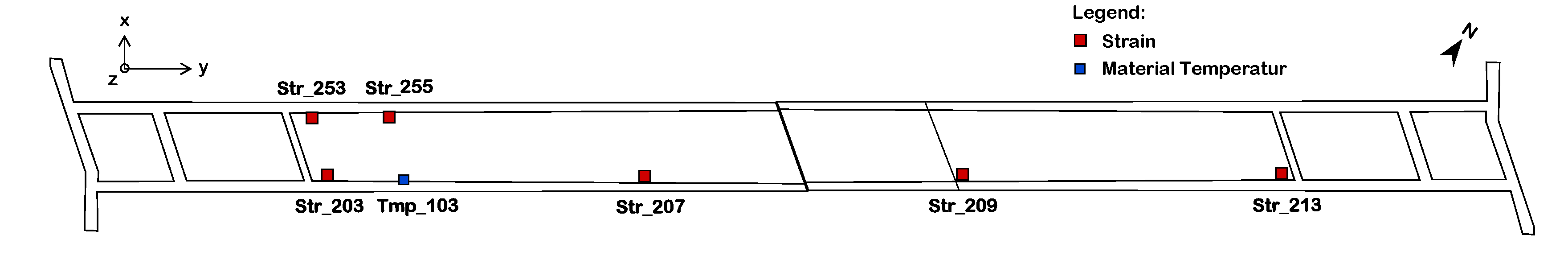}
    \caption{Vahrendorfer Stadtweg bridge view from the north (top-left). Three additional masses at the middle of the bridge (top-right). Overview of sensor positions (bottom).}
    \label{fig:vs}
\end{figure}

The prestressed concrete bridge ``Vahrendorfer Stadtweg'' is a 50-meter-long and 10-meter-wide structure constructed in 1972, see Figure~\ref{fig:vs}~(top left and middle). It spans the A7 freeway in the south of Hamburg, Germany, and features a single lane for agricultural traffic and a pedestrian walkway on the southeastern side. The bridge was built with an open frame design and a box girder cross-section.
The analysis presented here used measurements from six strain sensors (in the z-direction) and one material temperature sensor. The sampling frequency of 200~Hz resp.~10~Hz was downsampled (averaged) to one measurement per hour, compare \cite{Han.etal_2021}. The sensor placement is shown in Figure~\ref{fig:vs}~(bottom).
Load tests were performed on the bridge between February 22nd and March 23rd, 2024. These tests involved three large bags filled with sand, weighing 680~kg, 740~kg, and 740~kg, respectively. The bags were placed in the middle of the bridge. Each extra mass was placed for 10 days and gradually increased in steps to 680~kg (scenario A), 1,420~kg (scenario B), and 2,160~kg (scenario C). Figure~\ref{fig:vs}~(top right) shows the maximum mass (all three bags).
Afterward, the masses were positioned at the quarter point of the south side, and truck crossings were conducted; however, this second measurement period will not be included in the analysis reported here. For a more detailed description of the bridge and the load tests, see \cite{Koehncke.etal_2026}.

\section{Methodology}

\subsection{Conditional Principal Component Analysis and Feature Reconstruction}\label{sec_CC_FR}

In SHM, principal component analysis (PCA) is often employed for data normalization \cite{Reynders.etal_2014} and feature extraction \cite{Tibaduiza.etal_2016, Zhu.etal_2019}. The first principal components often capture EOVs such as temperature \cite{Cross.etal_2012}, whereas PCA features with smaller eigenvalues demonstrate robustness to environmental variations \cite{Kumar.etal_2020}.
So, the idea is to omit the leading principal components to remove the effects of EOVs on the data. One challenge, however, is the uncertainty about the number of components to exclude. Excluding too few may result in residual environmental effects, whereas excluding too many may lead to the loss of crucial information about damage. Moreover, it remains uncertain whether this implicit approach can completely mitigate the influence of EOVs, such as temperature variations \cite{Neumann.etal_2025, Neumann_2025}. Consequently, accounting for confounding variables more explicitly may be a more effective strategy. 

Standard PCA is an unsupervised method that can be performed via an eigendecomposition of the covariance matrix. But if instead of the marginal covariance matrix, a conditional covariance matrix $\boldsymbol{\Sigma}(z)$ is used, where the covariance varies with the confounding variable $z$, PCA becomes a supervised method. So, let $\mathbf{x} = (x_1,\ldots, x_p)^\top\in\mathbb{R}^{p}$ be a $p$-dimensional random (output) vector and $z\in\mathbb{R}$ a potential confounder, then the \emph{conditional} PCA can be carried out through \cite{Neumann.etal_2025}
\begin{equation}\label{eq_spd_cond}
	\boldsymbol{\Sigma}(z) = \textbf{A}(z)\boldsymbol{\Lambda}(z) \textbf{A}(z)^\top,
\end{equation}
with $\lambda_1(z),\ldots,\lambda_p(z)$ being the conditional eigenvalues and $\mathbf{a}_1(z),\ldots,\mathbf{a}_p(z)$ the conditional eigenvectors, which correspond to the (conditional) principal components. Note that the eigenvalues and eigenvectors can vary for different $z$, but conditional PCA requires the confounder to be measured. 
The conditional covariance matrix $\boldsymbol{\Sigma}(z)$ of $\mathbf{x}$ given a confounder $z$ used in Eq.~\eqref{eq_spd_cond} can be estimated through \cite{Yin.etal_2010, Neumann.etal_2025}
\begin{equation}
    \label{neumann:eq_Sdef}
	\hat{\boldsymbol{\Sigma}}(z; h) = \left\{\sum_{i=1}^n K_h(z_i - z)\left[\mathbf{x}_i - \hat{\mathbf{m}}(z_i)\right]\left[\mathbf{x}_i - \hat{\mathbf{m}}(z_i)\right]^\top\right\}\left\{\sum_{i=1}^n K_h(z_i - z)\right\}^{-1},
\end{equation}
where $\mathbf{x}_i = (x_{i1}, \ldots, x_{ip})^\top$, $i=1,\ldots,n$, are observations of $\mathbf{x}$, $z_i$ is the associated confounder variable (e.g., temperature), and $\hat{\mathbf{m}}(z_i)$ is an estimate of the mean of $\mathbf{x}$ at $z_i$. The conditional mean can, for instance, be estimated using methods such as penalized regression splines \cite{Eilers.Marx_1996, Neumann.Gertheiss_2022}, local polynomial regression \cite{Cleveland.etal_2017} or a Nadaraya-Watson kernel estimator \cite{Yin.etal_2010, Neumann.etal_2025}. 
$K_h(\cdot)$ is a kernel function with bandwidth $h$, with the latter being the smoothing parameter. The higher $h$, the wider the kernel and the smoother the estimated covariance function. 
The optimal bandwidth $h$ can be estimated using cross-validation, as, e.g., described in \cite{Yin.etal_2010, Petersen.etal_2019, Neumann.etal_2025, Neumann.etal_2025c}.

The eigenvalues and principal components from Eq.~\eqref{eq_spd_cond} can be used for \emph{feature reconstruction}. First, the corresponding scores \cite{Neumann.etal_2025}
\begin{equation}\label{eq_score_adjust}
	\textbf{s}_i = (\textbf{x}_i-\hat{\mathbf{m}}(z_i))^\top\textbf{A}(z_i) (\boldsymbol{\Lambda}(z_i))^{-1/2},
\end{equation}
with $(\boldsymbol{\Lambda}(z_i))^{-1/2} = \text{diag}(\lambda_1^{-1/2}(z_i),\ldots,\lambda^{-1/2}_p(z_i))$ are extracted. 
Using the (estimated) conditional mean $\hat{\mathbf{m}}(z_i)$, eigenvalues $\boldsymbol{\Lambda}(z_i)$, and principal components $\textbf{A}(z_i)$ removes the effect of the confounder $z$ from the component scores. These scores are uncorrelated, standardized quantities with a mean of zero and a variance of one for any given $z$-value~\cite{Neumann.etal_2025, Neumann_2025}. 

Then, the calculated scores $\mathbf{s}_i$ can be used to reconstruct the output data on the original, physically meaningful scale, with the confounder effect removed. Specifically, we have \cite{Neumann.etal_2025c}
\begin{equation}\label{eq_reconFeat}
    \tilde{\textbf{x}}_i= \bar{\mathbf{x}} +  \textbf{A}(\boldsymbol{\Lambda})^{1/2}\textbf{s}_i^\top,
\end{equation}
with $\bar{\mathbf{x}}$ the marginal mean of $\mathbf{x}$, and $(\boldsymbol{\Lambda})^{1/2} = \text{diag}(\lambda_1^{1/2},\ldots,\lambda_p^{1/2})$ the eigenvalues and $\textbf{A} = [\mathbf{a}_1 \ldots \mathbf{a}_p]$ the principal components from common PCA of the residual data, i.e., the (estimated) conditional mean given the confounder values had been subtracted from the measurements. The covariance of the residuals is also known under the name \emph{partial} covariance.
The marginal mean $\bar{\mathbf{x}}$ and partial eigenvalues and principal components are used for back-transforming instead of the conditional ones to prevent re-inducing the confounder effect removed in Eq.~\eqref{eq_score_adjust}. Standard marginal PCA is not used in Eq.~\eqref{eq_reconFeat} because the first principal component(s) mainly account(s) for EOV(s), resulting in larger leading eigenvalues and over-weighting of the first component(s), which increases variation in the reconstructed output.

\subsection{Control Charts}\label{sec:control_charts}

The reconstructed features can be monitored using an appropriate control chart.
Here, a multivariate memory-type control chart, the ``Multivariate Exponentially Weighted Moving Average'' (MEWMA) \cite{Lowry.etal_1992} is used to monitor the reconstructed output data from Eq.~\eqref{eq_reconFeat}, under the assumption that $\tilde{\mathbf{x}}_i \sim \mathcal{N}(\boldsymbol{\mu},\boldsymbol{\Gamma})$ .
This type of control chart also incorporates past information, thereby increasing its sensitivity to small shifts. 

First, a mean vector is defined based on a change point model, specifically, \(\boldsymbol{\mu} = \boldsymbol{\mu}_0\) if \(i < \vartheta\) and \(\boldsymbol{\mu} = \boldsymbol{\mu}_1\) if \(i \geq \vartheta\) for an unknown time point \(\vartheta\)~\cite{Knoth_2017}.
Then, with smoothing parameter $0 < \kappa \leq 1$, a smoothing procedure is applied to estimate the MEWMA statistic
\begin{equation}\label{eq:MEWMA}
 \boldsymbol{\omega}_i=(1-\kappa)\boldsymbol{\omega}_{i-1} + \kappa\tilde{\mathbf{x}}_i, \quad  \boldsymbol{\omega}_0=\boldsymbol{0},\
\end{equation}
for time point $i = 1, 2, \ldots$ . The smoothing parameter $\kappa$ controls the sensitivity of the shift to be detected, i.e., smaller values such as $\kappa \in \{0.1,0.2,0.3\}$ are usually selected to detect smaller shifts \cite{Hunter_1986}, whereas $\kappa = 1$ results in the Hotelling chart \cite{Hotelling_1947}. The control statistic is the Mahalanobis distance
\begin{equation}\label{eq:MHD}
T^2_i=(\boldsymbol{\omega}_i-\boldsymbol{\mu}_0)^\top\boldsymbol{\Gamma}^{-1}_{\boldsymbol{\omega}}   (\boldsymbol{\omega}_i-\boldsymbol{\mu}_0),
\end{equation}
with the asymptotic covariance matrix $\boldsymbol{\Gamma}_{\boldsymbol{\omega}}$ of $\boldsymbol{\omega}_i$, $\boldsymbol{\Gamma}_{\boldsymbol{\omega}} = \lim_{i\to\infty} \text{Cov}(\boldsymbol{\omega}_i) = \left(\frac{\kappa}{2-\kappa}\right)\boldsymbol{\Gamma}$.

The MEWMA control chart triggers an alarm if the control statistic $T^2_i$ exceeds the threshold value $h_4$. On in-control data, the statistic \eqref{eq:MEWMA} is typically reinitialized after a (false) alarm. ``In-control'' means that only common causes, not special causes like damage, influence the system, while ``out-of-control'' indicates the presence of special causes such as damage. The in-control ``average run length'' ARL$_0$ is used to calibrate the threshold $h_4$. In general, the ARL is defined as the expected number of observations until the threshold is exceeded. 
Therefore, the in-control ARL$_0$ should be high to minimize false alarms, whereas if the process is out-of-control, the ARL (the so-called ARL$_1$) should be low to detect changes quickly. 
Since the reconstructed outputs are not independent and identically distributed (i.i.d.) but correlated over time, a simulation-based method is used for calibration. It estimates the ARL by dividing the data into daily blocks and resampling with replacement to create new datasets. In the present study, the ARL is calculated using Eq.~\eqref{eq:MEWMA} and \eqref{eq:MHD} with 100,000 repetitions, and the control limit is chosen to achieve an ARL$_0$ of 720 (hours), i.e., 30~days. The corresponding algorithm is presented in more detail in \cite{Neumann.etal_2025c}. To distinguish the training phase and out-of-sample monitoring, the terms ``Phase~I'' and ``Phase~II'' will be used from now on. Phase~I consists only of in-control data, which can be used for model training, whereas Phase~II may consist of both in-control and out-of-control data.

\section{Application to SHM Load-Test Data}\label{AppSHMData}

The strain measurements from July 2nd, 2023, to February 17th, 2024, are used as Phase~I data to estimate the conditional mean and covariance. The time between 
February 18th and March 23rd, 2024, will be considered Phase II; this covers the first three additional mass scenarios described above. This data will also be used to calculate the control chart to see whether ``damage'' (i.e., the additional masses) can be detected. The conditional mean was estimated using penalized regression splines \cite{Eilers.Marx_1996, Neumann.Gertheiss_2022}. The optimal bandwidths for the conditional covariance matrix were initially estimated via cross-validation, as described in \cite{Neumann.etal_2025}, resulting in small values between $0.1$ and $0.3$. The conditional covariance was estimated separately for each sensor pair, cf. Eq.~\eqref{neumann:eq_Sdef}. Due to the sparse data between $10^\circ$C and $15^\circ$C, additional smoothing was necessary. To achieve this, the bandwidth was increased to $1.2$ to widen the kernel. A wider kernel allows for more data points from adjacent temperatures to be included in the analysis, stabilizing the results. In what follows, the results for three different methods for feature reconstruction will be shown, each utilizing a different type of temperature adjustment: 

\begin{itemize}
    \item[(a)] \textit{Unsupervised}: The unsupervised method estimates the common PCA on the measurements, but omits the first principal component when estimating the scores and, subsequently, the reconstructed features. The first principal component is excluded from the estimation because it is assumed to primarily account for operational and environmental effects \cite{Cross.etal_2012}. This method is unsupervised as it does not need the confounding variable to be measured.

    \item[(b)] \textit{Partial}: For the partial method, the conditional mean is estimated using penalized regression splines \cite{Eilers.Marx_1996,Neumann.Gertheiss_2022}. Then the conditional mean is subtracted from the measurement, i.e., the residuals are estimated, and the marginal mean is added to vertically shift the data to the original scale. Using the conditional mean transforms the partial method into a supervised method, as it requires the confounder to be measured. 
    
    \item[(c)] \textit{Conditional}: The conditional method as discussed in Section~\ref{sec_CC_FR} is applied utilizing the conditional mean and the conditional covariance to remove the temperature influence in the mean \textit{and} in the covariance. As with the previous method, the conditional method is a supervised method that requires the confounder to be measured.
\end{itemize}

\subsection{Reconstructed Strain Data}\label{sec:results_vs}

\begin{figure}[!h]
    \centering
    \includegraphics[width=\linewidth]{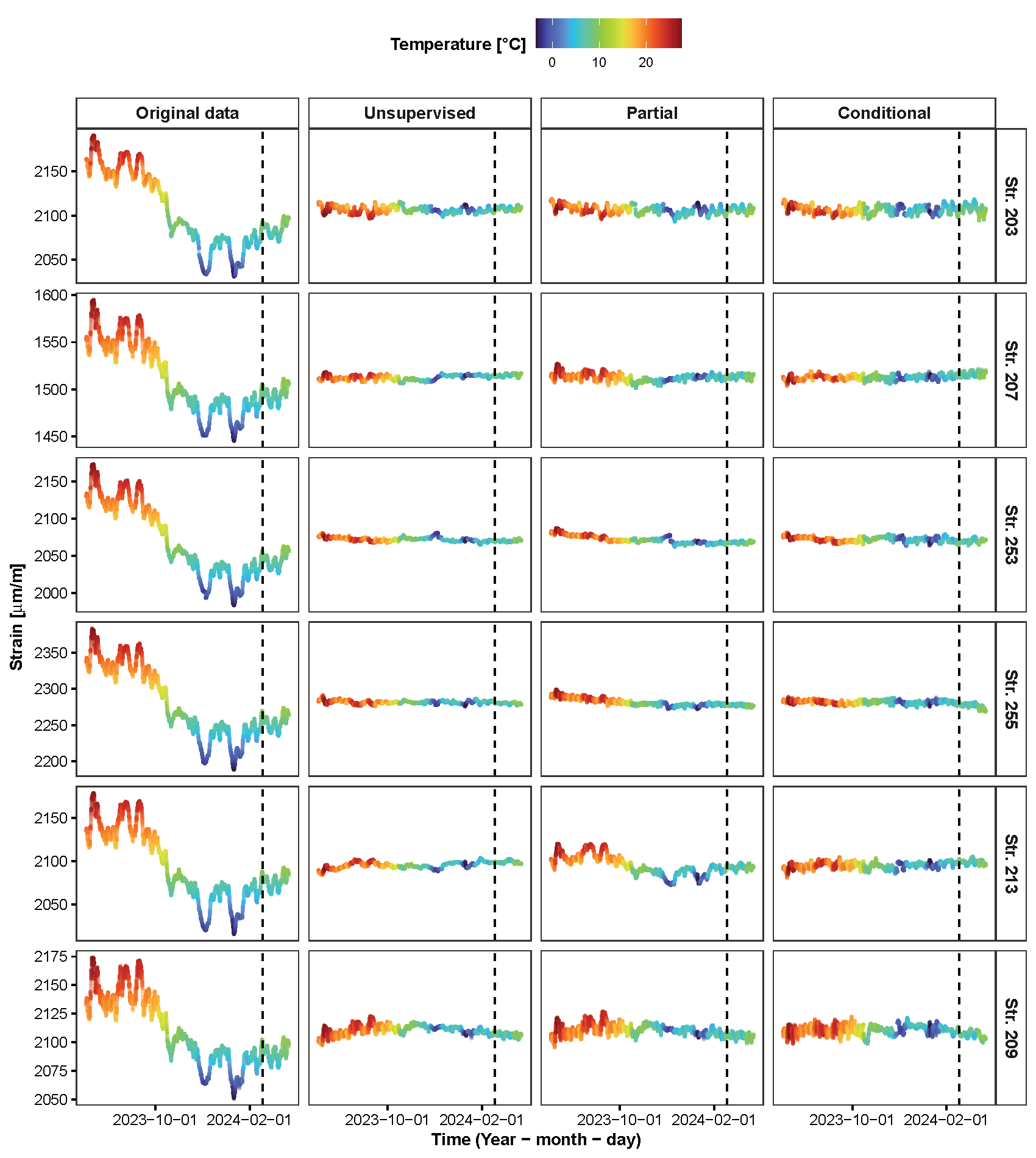}
    \caption{Original strain data (left) and the reconstructed strain data using the unsupervised (center left), partial (center right), and conditional (right) method of the Vahrendorfer Stadtweg. The vertical dashed line separates Phase I from Phase II. The data is colored according to the measured temperature.}
    \label{fig:rf_vs_time}
\end{figure}

Figure~\ref{fig:rf_vs_time} shows the original data over time in the first column. The rows correspond to the six strain sensors from Figure~\ref{fig:vs}. 
The results of unsupervised feature reconstruction using principal components two to six are found in the second column, the results of the approach using partial PCA in the third column, and the proposed conditional method in the fourth column. For the latter two, all principal components are used. Consequently, the third column basically shows the residuals obtained by subtracting the estimated conditional mean from the observed data (the first column). The data is colored according to the measured temperature at each time point, i.e., from blue (cold) to red (warm). 
In the original data, the strain measurements and temperature are highly correlated. This is no longer the case with the reconstructed data; most of the reconstructed outputs appear stable over time and temperature, except for strain sensor 213. Using the partial method, the dependency could not be completely eliminated, although the association is less pronounced than for the original data. 
All three methods demonstrate the clear advantage of feature reconstruction. By using either (a), (b), or (c) from above, reconstructed data are obtained that align with the original measurements but have been adjusted for temperature. This enables the reconstruction of physically meaningful features without the oscillations caused by EOVs in the original data.

\begin{figure}[!h]
    \centering
    \includegraphics[width=\linewidth]{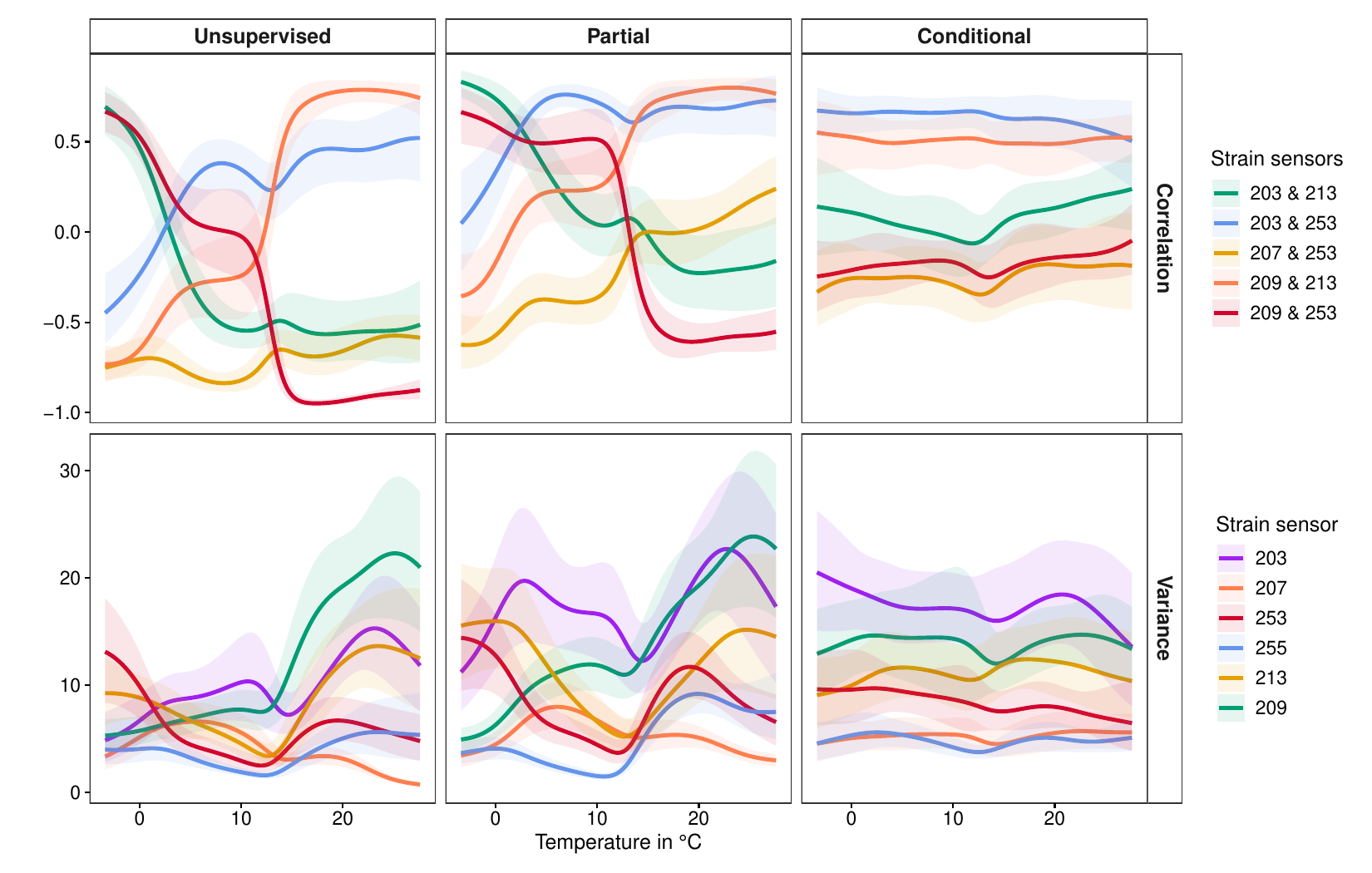}
    \caption{Conditional correlations (top) and variances (bottom) as functions of temperature, together with approximate, pointwise $95\%$ confidence intervals, of the reconstructed strain data from the Vahrendorfer Stadtweg if using the unsupervised (left), partial (middle), or conditional (right) approach.} 
    \label{fig:rfcv_vs}
\end{figure}

However, Figure~\ref{fig:rfcv_vs} shows the conditional correlation (top, solid lines) and variance (bottom, solid lines) for a selection of sensor pairs, plotted against temperature, together with the approximate, pointwise $95\%$ confidence intervals (shaded). 
The confidence interval was calculated using Eq.~\eqref{neumann:eq_Sdef} and block bootstrapping, as in \cite{Neumann.etal_2025b,Neumann.etal_2025c}. Block bootstrapping is used to resample the data by dividing it into blocks of days, followed by estimating the conditional covariance~\eqref{neumann:eq_Sdef}. This is repeated $10^5$ times to ensure that a comprehensive range of variations is captured. The approximate, pointwise $95\%$ confidence interval is then calculated from the repetitions. A more detailed description is given in \cite{Neumann.etal_2025b, Neumann.etal_2025c}.
If using the partial method, the temperature dependence is adjusted only in the mean, so the conditional correlations and variances correspond to those of the original measurements; therefore, only the results of the partial method are shown (not for the original data, as they are identical). 
Those are shown in the second column and vary with different temperatures, indicating that temperature still influences the (co)variance. 
Omitting the first principal component, i.e., the unsupervised method (first column), results in a slight improvement regarding the variance but has little to no effect on the correlation. However, the conditional correlations and variances of the reconstructed data remain nearly constant when the newly proposed conditional approach is used, demonstrating that it can eliminate the temperature effect not only in the mean but also in the (co)variance. 
The confidence intervals in Figure~\ref{fig:rfcv_vs} show that most of the estimates are stable across bootstrap samples, but increased uncertainty exists at the boundaries and around $10^\circ$C due to sparse data.

\subsection{Monitoring Results}\label{sec:monitoring_vs}

\begin{figure}[!h]
    \centering
    \includegraphics[width=\linewidth]{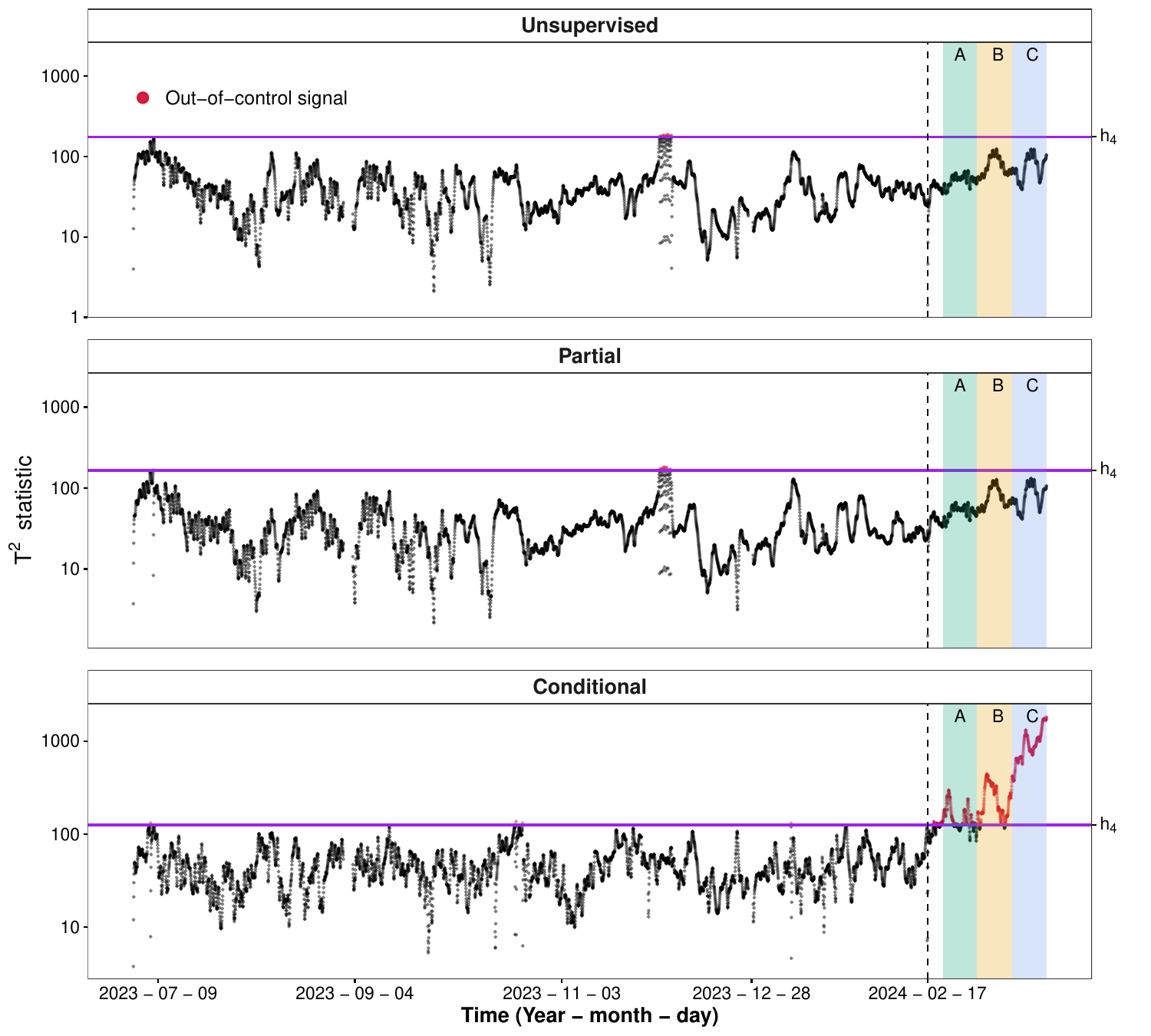}
    \caption{MEWMA control chart (with $\kappa = 0.2$) on log scale for the unsupervised ($h_4 = 175.93$), partial ($h_4 = 165.90$), and conditional ($h_4 = 125.00$) approach for feature reconstruction for the Vahrendorfer Stadtweg strain data. The control limits $h_4$ are calculated using block bootstrapping. The dashed line separates Phase~I from Phase~II, with the three load test scenarios highlighted in color.}
    \label{fig:MEWMA_vs}
\end{figure}

The MEWMA control chart (on log scale) from Eq.~\eqref{eq:MEWMA} and \eqref{eq:MHD} is shown in Figure~\ref{fig:MEWMA_vs} for smoothing parameter $\kappa = 0.2$ and control limits (purple horizontal lines) $h_4 = 175.93$ for the unsupervised, $h_4 = 165.90$ for the partial, and $h_4 = 125.00$ for the conditional approach. The control limit was estimated using block bootstrapping with the reconstructed data being divided into blocks of days, $10^5$ repetitions, and an ARL of $1,000$. 
The values above the control limit are plotted in red. The dashed vertical line indicates the beginning of Phase~II, and the shaded areas highlight the three additional mass scenarios (light to heavy). 

Two periods of false alarms stand out. For the unsupervised and partial method, there are false alarms at the beginning of December, when temperatures were low (below $5^\circ$C). Here, the control statistic of the conditional method did not trigger the alarm, but in mid-October, when the temperature was around $10^\circ$C, it did. The latter false alarms could be due to the sparse training data and, thus, due to an inaccurate estimation of the conditional covariances.

For both the unsupervised and partial methods, the control statistic does not exceed the control limit at any time. Therefore, the unsupervised and partial methods failed to detect the additional masses. By contrast, if using the conditional method, the control statistic is almost always above the control limit, with its height varying across scenarios. The control statistic is lower for the first two additional mass scenarios and higher for the third one. The spike in the control statistic at the beginning of the first scenario can be explained by a car with a trailer and a wheel loader crossing the bridge, which had to be used for repositioning the additional mass.

Table~\ref{tab:false_alarms_bb} summarizes the false alarms and probability of detection for each of the three methods, along with different values of parameter $\kappa$. The respective control limits were calculated using block bootstrapping, but the results are similar to those of moving-block bootstrapping. 
As mentioned in Section~\ref{sec:control_charts}, the parameter $\kappa$ controls the sensitivity of the shift to be detected. Thus, the values $\kappa = 0.1, 0.3, 1$ were added for comparison. 
As can be seen again, neither the partial nor the unsupervised methods can detect the additional masses.

\begin{table}[t]
    \centering
    \tabcolsep7.1pt 
    \small
    \caption{False alarms and probability of detection (POD) per method and varying smoothing parameter $\kappa$ for the bridge Vahrendorfer Stadtweg. The respective control limits $h_4$ were calculated using block bootstrapping.}
    \begin{tabular}{l *{9}{r}}
	\toprule
        \multicolumn{1}{c}{} & 
	\multicolumn{3}{c}{\textbf{Unsupervised}} &  
        \multicolumn{3}{c}{\textbf{Partial}} & 
        \multicolumn{3}{c}{\textbf{Conditional}} \\
        \multicolumn{1}{c}{\textbf{$\kappa$}} &
	\multicolumn{2}{c}{False Alarms} &  
        \multicolumn{1}{c}{POD} & 
	\multicolumn{2}{c}{False Alarms} &  
        \multicolumn{1}{c}{POD} &
	\multicolumn{2}{c}{False Alarms} &  
        \multicolumn{1}{c}{POD}\\
	\midrule
        $0.1$ & $\textbf{7}$ & $[\textbf{0.13}\mathbf{\%}]$ & $0\%$ & $9$ & $[0.17\%]$ & $0\%$ & $12$ & $[0.22\%]$ & $\textbf{100}\%$  \\
	\midrule
        $0.2$ &  $9$ & $[0.17\%]$ & $0\%$ & $11$ & $[0.20\%]$ & $0\%$ & $\textbf{6}$ & $[\textbf{0.11}\mathbf{\%}]$ & $\textbf{87.4}\mathbf{\%}$  \\
	\midrule
        $0.3$ &  $12$ & $[0.22\%]$ & $0\%$ & $15$ & $[0.28\%]$ & $0\%$ & $\textbf{6}$ & $[\textbf{0.11}\mathbf{\%}]$ & $\textbf{75.73}\mathbf{\%}$  \\
	\midrule
        $1.0$ & $85$ & $[1.57\%]$ & $0\%$ & $92$ & $[1.70\%]$ & $0\%$ & $\textbf{21}$ & $[\textbf{0.39}\mathbf{\%}]$ & $\textbf{64.08}\mathbf{\%}$  \\
	\bottomrule
    \end{tabular}
    \label{tab:false_alarms_bb}
\end{table}
%

\section{Summary and Conclusion}\label{conclusion}

This paper presented a method for calculating confounder-adjusted features from sensor measurements. The reconstructed features are in the same domain as the original measurements, enabling physical interpretation. The presented new approach uses conditional covariances and subsequent (conditional) principal component analysis (PCA) to ensure reliable results and insights. In contrast to standard, unsupervised PCA, the new method requires measuring confounding variables, such as environmental and operational variables (EOVs), thus transforming PCA from an unsupervised to a supervised method. 
Considering a real-world bridge dataset, the reconstructed features showed no dependence on temperature in either the mean or the (co-)variance anymore. Existing approaches for confounder removal (e.g., omitting the first principal component or response surface modeling), by contrast, could not fully eliminate the influence of the confounding variable. This was especially noticeable in the conditional (co-)variances. While temperature dependencies in the mean could be removed, covariances still varied (non-linearly) with temperature.

Limitations of the new, conditional approach include the need to measure confounding variables both during the in-control phase to ensure reliable training, as demonstrated with the Vahrendorfer Stadtweg bridge (where infrequent temperature measurements led to unreliable or no reconstructed features) and during the entire monitoring phase. 

Additionally, an advanced multivariate control chart, the MEWMA control chart, was established to monitor the reconstructed features and determine whether damage could be detected. To evaluate the performance of the considered methods, two factors were examined: the false alarm rate and the probability of detection. The false alarms could be reduced using the conditional method. More substantially, however, the probability of detection could be increased to $64\%$-$100\%$ when using the conditional approach, compared to $0\%$ for the unsupervised and partial methods. In other words, only the conditional approach detected the additional masses, especially the lightest, whereas the unsupervised and partial methods failed to detect even the heaviest mass. 

\section*{Code Availability}

Code implementing the output reconstruction and the monitoring procedure is available on GitHub at \url{https://github.com/neumannLizzie/FeatureReconstructionSHM/}.

\section*{Acknowledgements}
This research paper out of the project ``SHM -- Digitalisierung und Überwachung von Infrastrukturbauwerken'' is funded by dtec.bw -- Digitalization and Technology Research Center of the Bundeswehr, which we gratefully acknowledge. dtec.bw is funded by the European Union -- NextGenerationEU.

\bibliographystyle{unsrtnat}
\bibliography{covest.bib}

\end{document}